\documentclass[11pt, oneside]{article}
\usepackage{geometry}
\usepackage{graphicx}
\usepackage{amssymb}
\usepackage{amsmath}
\usepackage{url}
\usepackage{adjustbox}
\usepackage{comment}
\usepackage{caption}
\usepackage{subcaption}
\usepackage{booktabs}
\usepackage{tabularx}
\usepackage{array}
\usepackage{authblk}
\usepackage{setspace}
\usepackage{breakcites}
\usepackage{float}
\usepackage{xspace}
\usepackage[authoryear, round]{natbib}
\usepackage{booktabs}
\usepackage{lineno}
\usepackage[nobiblatex]{xurl}

\makeatletter

\title{Statistical Issues in the Diagnosis of Shaken Baby Syndrome/Abusive Head Trauma}

 \author[1,2]{Maria Cuellar}
 
 \affil[1]{Department of Criminology, University of Pennsylvania, 3718 Locust Walk, Philadelphia, PA, 19104, United States}
 
 \affil[2]{Department of Statistics and Data Science, Wharton School, University of Pennsylvania, Walnut Street, Philadelphia, PA 19104, United States}

\begin{document}

\maketitle

\begin{abstract} 
The diagnosis of Shaken Baby Syndrome/Abusive Head Trauma (SBS/AHT) is fraught with controversy due to critical statistical deficiencies in the data underpinning these diagnoses. This paper examines the reliability and scientific foundation of SBS/AHT through a statistical lens, highlighting the lack of independently verified ground truth, contextual biases, data circularity, and diagnostic heterogeneity. These issues render current methodologies inadequate and complicate evaluations of diagnostic accuracy, particularly when legal determinations are integrated into medical assessments. Without empirical evidence validating the specificity of symptoms like subdural hematoma, retinal hemorrhage, and brain swelling, the diagnosis remains untested and its foundational validity unproven. We recommend that physicians focus on reporting observed clinical signs and avoid making determinations of abuse, which should remain within the legal domain. Addressing these challenges requires comprehensive, high-quality data collection encompassing contextual, medical, and legal information to evaluate the accuracy, repeatability, and reproducibility of SBS/AHT diagnoses. These efforts are essential to protect vulnerable children while ensuring fairness and accuracy in legal proceedings involving allegations of abuse.
\end{abstract}


\section{Introduction}

In September, 2024, the Texas Court of Criminal Appeals ruled that the Shaken Baby hypothesis, used to convict Andrew Roark of Dallas, lacks scientific validity and granted him a new trial. This same hypothesis, supported by nearly identical testimony from the same child abuse specialist, was central to Robert Roberson's 2003 conviction in Anderson County, Texas. Despite compelling evidence that Roberson's daughter's tragic death resulted from natural causes, he has not been granted a new trial. This disparity has reignited controversy, particularly as Roberson remains on death row following a delayed execution. Critics, including the Innocence Project, argue that Shaken Baby Syndrome is a debunked hypothesis, while the American Academy of Pediatrics maintains that it is a serious and ``clearly definable'' form of child abuse. The contrasting outcomes in these cases highlight the broader debate over the reliability of this diagnosis in the justice system. 

But how should a physician, or a pathologist, determine whether the child's brain condition was caused by an intentional act? Some researchers say that there is often no scientifically valid way to answer this question in general \citep{findley2023shaken}, and others say that it can be determined by doing patient observation, just like with any other medical diagnosis \citep{parks2012pediatric}. For researchers without a medical or legal opinion on the matter of SBS/AHT, the opposing claims in the controversy \citep{findley2019feigned} are difficult to evaluate due to the lack of high-quality data, both regarding medical and legal information. 

The American Academy of Pediatrics and the CDC have attempted to standardize the definition and diagnosis of AHT \citep{aapreport} For example, the CDC defines the diagnosis as follows \citep{parks2012pediatric},
\begin{quote}
    Pediatric abusive head trauma is defined as an injury to the skull or intracranial contents of an infant or young child ($<$ 5 years of age) due to inflicted blunt impact and/or violent shaking.
\end{quote} 
However, disagreements persist. Despite extensive research efforts, determining the prevalence of the  diagnosis of SBS/AHT accurately remains challenging. The scientific foundations of Shaken Baby Syndrome/Abusive Head Trauma (SBS/AHT) require thorough investigation, as a diagnosis of SBS/AHT often implies a criminal conviction.

This article highlights the need for a systematic effort to collect and analyze data related to SBS/AHT, to standardize diagnostic practices, and to ensure that legal decisions are based on reliable and objective medical evidence. This is done from a statistical perspective. Addressing these issues is essential to safeguard the rights of accused individuals and to protect vulnerable children from abuse.

This article does not claim that SBS/AHT is ``junk science'' or a ``fake'' diagnosis. Instead, it claims that given that the data and statistical analyses that exist are of low quality, there is no evidence that the medical diagnosis is indeed a medical diagnosis separate from other brain conditions. It is possible that the diagnosis is simply a different brain condition in conjunction with contextual information from parties (such as the police, multidisciplinary teams, or court decisions). However, we will not know that this is the case until we have better data describing the actual diagnostic procedure. With more comprehensive data about the contextual information, and about who made what claim, we can begin to understand the mechanisms of the condition, the diagnosis, and the legal decisions, separately.

\section{The data about SBS/AHT diagnoses is insufficient to understand the diagnosis}

The diagnosis of Shaken Baby Syndrome/Abusive Head Trauma (SBS/AHT) lacks high-quality data due to several significant limitations. Some of these points have been covered by others \citep{findley2011shaken} from a legal perspective. This article covers them from a statistical perspective. Table \ref{tab:flaws} summarizes the flaws in the data, and the following sections expand on each point.

\begin{table}[!ht] 
\centering 
\begin{tabular}{p{0.45\linewidth} p{0.45\linewidth}} 
\toprule
\textbf{Flaw} & \textbf{Impact} \\ 
\midrule
\textbf{1. There is no ground truth}: There is no independently verified source of information about whether a child was abused. &
No statistical model can be used to predict abuse accurately without additional information. \\ 
\textbf{2. There is circularity in the data}: Legal results from previous cases are used to inform whether future medical diagnoses are considered abuse. & 
Incorrect determinations of abuse (from the law) can be taken as correct, and this can create a feedback loop that reinforces incorrect determinations. \\ 
\textbf{3. There is contextual bias}: Since the determination about whether a child was abused is subjective, it is prone to bias due to irrelevant contextual information. & 
This can lead to incorrect determinations of abuse. \\ 
\textbf{4. There is heterogeneity across physicians}: Due to subjectivity in the diagnosis and variability in the contextual information, physicians do not agree about which cases were caused by abuse. & 
Since the medical determination of abuse is subjective and based on different types of contextual information, there is variability in diagnosing, and the correct diagnosis is never confirmed by ground truth. \\  
\textbf{5. Medical and legal decisions are combined}: Physicians and pathologists opine about whether a child was abused, but this should be left to the trier of fact because it is a legal determination. & 
It is improper to have a physician make legal decisions. And this creates a feedback loop from legal proceedings to future medical diagnoses, which is also improper and has unknown error rates. \\ 
\textbf{6. Fatal cases have additional complications}: Pathologists become involved in fatal cases, and they learn the contextual information as well. & 
The accuracy of the additional decision about cause and manner of death is unknown. \\ 
\bottomrule
\end{tabular}
\caption{An example of a table with two columns, where the left column is wider than the right.} 
\label{tab:flaws}
\end{table}

\subsection{There is no ground truth}

First, and most importantly, there is no ground truth about whether a child was really abused. Ground truth refers to information verified as true or real, gathered through direct observation or measurement (i.e., empirical evidence), rather than derived from inference. There \textit{could be} ground truth, if there was evidence that a child was shaken or abused (e.g., in a video) and this resulted in the brain conditions observed in SBS/AHT cases. However, this has never been independently witnessed \citep{chp3neuropathology}. Infants and young children cannot testify reliably, and any witness accounts may be unreliable or biased. In most forensic disciplines, ground truth is crucial for validation, yet SBS/AHT diagnosis lacks this essential component, making it challenging to assess diagnostic accuracy. Since we do not have a ground truth, could it be that some cases of sudden infant death syndrome incorrectly diagnosed as shaken baby syndrome? \cite{chp7diagnosis} addresses this question specifically.

\subsection{There is circularity in the data}

\begin{figure}[!ht]
   \centering
   \includegraphics[width=.6\textwidth]{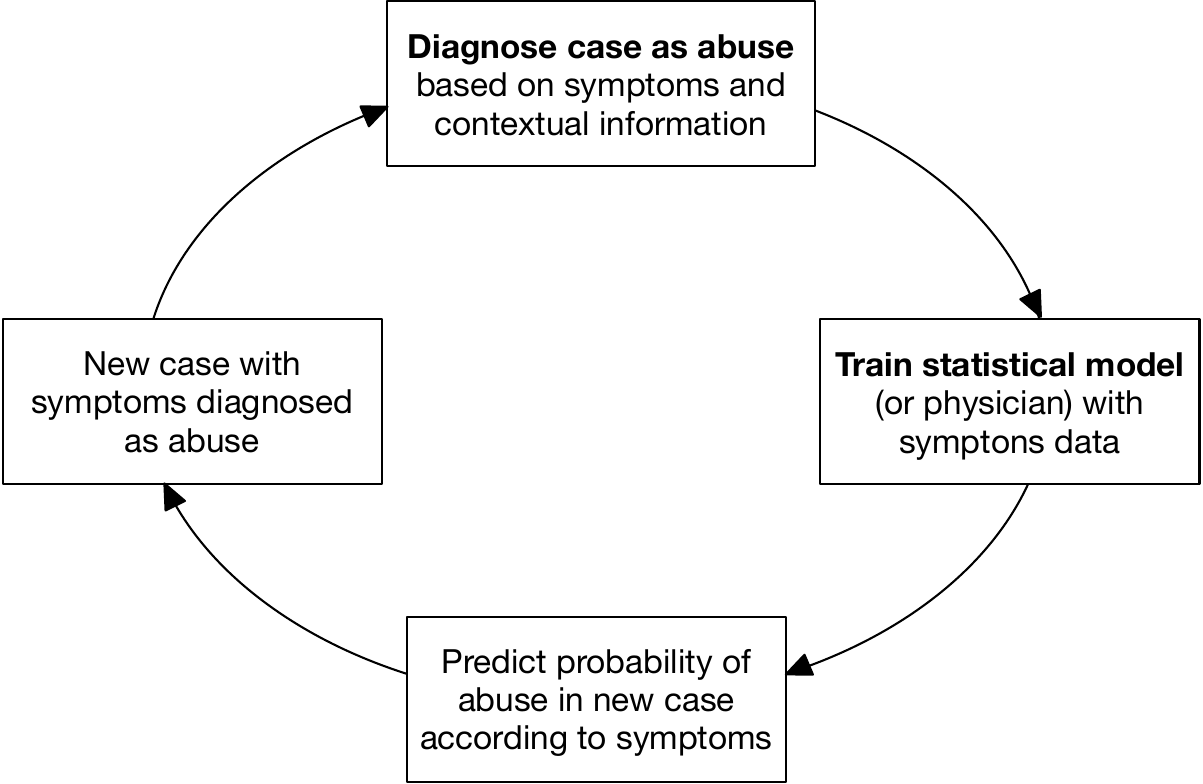} 
   \caption{Flowchart depicting the circularity in data and statistical models about SBS/AHT. The additional contextual information used to make diagnoses is not included here, but it likely plays a role in real casework.}
   \label{fig:circularity}
\end{figure}

Circularity is a significant concern with SBS/AHT data because diagnoses often rely on self-confirming criteria rather than objective, independently verified evidence. Figure \ref{fig:circularity} shows the circular reasoning that occurs in SBS/AHT diagnoses. In SBS/AHT cases, diagnostic conclusions are frequently drawn from a specific combination of medical findings--typically subdural hemorrhage, retinal hemorrhage, and brain swelling--that has become established as ``diagnostic'' of abuse. 

Circular reasoning occurs when clinicians interpret these findings as proof of abuse without independent evidence of how the injuries occurred. As a result, these diagnostic criteria become self-validating, where the presence of these findings leads to an assumption of abuse, and confirmed cases are then used to reinforce the assumption that the findings indicate abuse \citep{chp17causal}.

Studies that use diagnosed SBS/AHT cases to validate these findings perpetuate the cycle, as these cases were originally classified based on the same unverified criteria. One benefit of having statistical models or algorithms, though, is that they clarify reasoning that usually happens qualitatively in humans' minds, without explanation. In \cite{maguire2}, which used the results of \cite{maguire1} and was analyzed in \cite{cuellar2017causal}, the data used to train a model was obtained from only six physicians, without a discussion of what those physicians' perspectives about SBS/AHT are. 

Furthermore, and crucially, the contextual information used by these physicians is not contained in the training data, and thus the model cannot repeat the actions of making a diagnosis because the data is incomplete. This circularity undermines the reliability of SBS/AHT data, as it can lead to biased interpretations and the reinforcement of potentially flawed diagnostic criteria without independent corroboration.
Studies, such as \cite{maguire2}, which was analyzed in \cite{cuellar2017causal}, that use diagnosed SBS/AHT cases to validate these findings perpetuate the cycle, as these cases were originally classified based on the same unverified criteria. In \cite{maguire2}, the data used to train a model was obtained from only six physicians, without a discussion of what those physicians' perspectives about SBS/AHT are. 

A review of prior literature further demonstrates the circularity. Articles claiming that SBS/AHT is defined by certain symptoms reference prior articles as the argument for why these are the correct symptoms to select. Those articles reference prior articles, and so on, without there ever being an external verification that those symptoms are indeed caused by abuse. For instance, \cite{paine2019prevalence}'s attempt to justify their use of the data by citing previous literature, and state that rib fractures in young children ``have high specificity for child abuse.'' But when one traces these citations all the way to the initial literature on this topic, one finds that there is no ground truth there, either. The initial claim \citep{guthkelch1971} was an assumption, which has subsequently been taken as a fact \citep{guthkelch2011problems}. \cite{paine2019prevalence}'s citation of previous medical literature as a justification, since that cited literature assumes that abuse is the cause of these features, without establishing ground truth through external validation, does not eliminate circular reasoning but perpetuates it.

This problem is not remediated by the CDC's workshop gathering physicians to select diagnoses and external codes for fatal and non-fatal cases of SBS/AHT because the physicians might have used the literature, and might have been trained by other physicians who read this literature, without there being an external verification of the symptoms.

\cite{maguire2} tries to avoid this circularity by allowing for more information to be included in the decision of whether a child was abused. They provide a ``ranking criteria'' showing how abuse was determined for the children in their data: ``(1) Abuse confirmed at case conference or civil, family, or criminal court proceedings or admitted by perpetrator or independently witnessed; (2) Abuse confirmed by stated criteria, including multidisciplinary assessment; (3) Diagnosis of abuse defined by stated criteria; (4) Abuse stated as occurring, but no supporting detail given as to how it was determined; (5) Abuse stated simply as ``suspected''; no details on whether it was confirmed.'' These all have possible errors and do not provide a ground truth.

\subsection{There is contextual bias}\label{sec:contextualbias}

Diagnosis often depends on the physician's ``gut feeling,'' which is influenced by experience and subjective observations, such as interpreting a caregiver's behavior. For example, a mother checking her watch might be taken as a sign of disinterest, affecting the physician's judgment. To verify diagnoses, physicians sometimes review cases in monthly meetings, where they reach a consensus, yet this approach only compounds the reliance on subjective judgment. Contextual bias also poses a challenge (\cite{cuellar2022probabilistic}, \cite{chp13contextualbias}), as irrelevant details like caregiver demeanor can inadvertently influence diagnostic conclusions, potentially leading to erroneous findings. 

Researchers found that cognitive bias affected forensic pathologists' decision-making \cite{dror2021cognitive}. The pathologists were more likely to declare a case a homicide than an accident when the child was Black and the caretaker was the mother's ex-boyfriend relative to when the child was white and the caretaker was the child's grandmother. This research has received lots of criticism and responses from the pathology community\footnote{See: \url{https://onlinelibrary.wiley.com/doi/10.1111/1556-4029.14697}. Accessed: 11-19-2024.} Nevertheless, it is evidence that the determination of abuse is subject to bias from extraneous information.

\subsection{There is heterogeneity across physicians}

Variation in clinical practices makes diagnosing SBS/AHT accurately challenging. There is significant heterogeneity in SBS/AHT diagnoses, which vary by state, hospital, and type of facility, further undermining diagnostic consistency and reliability. Together, these factors underscore the lack of a standardized approach to SBS/AHT. Physicians rarely witness both the clinical features and the abuse, making it difficult to establish a ground truth a posteriori. Instead, they infer abuse from interviews with family members, observations by paramedics, or caretaker behaviors, all pieces of information that might vary greatly from case to case.

\begin{figure}[!ht]
   \centering
   \includegraphics[width=.8\textwidth]{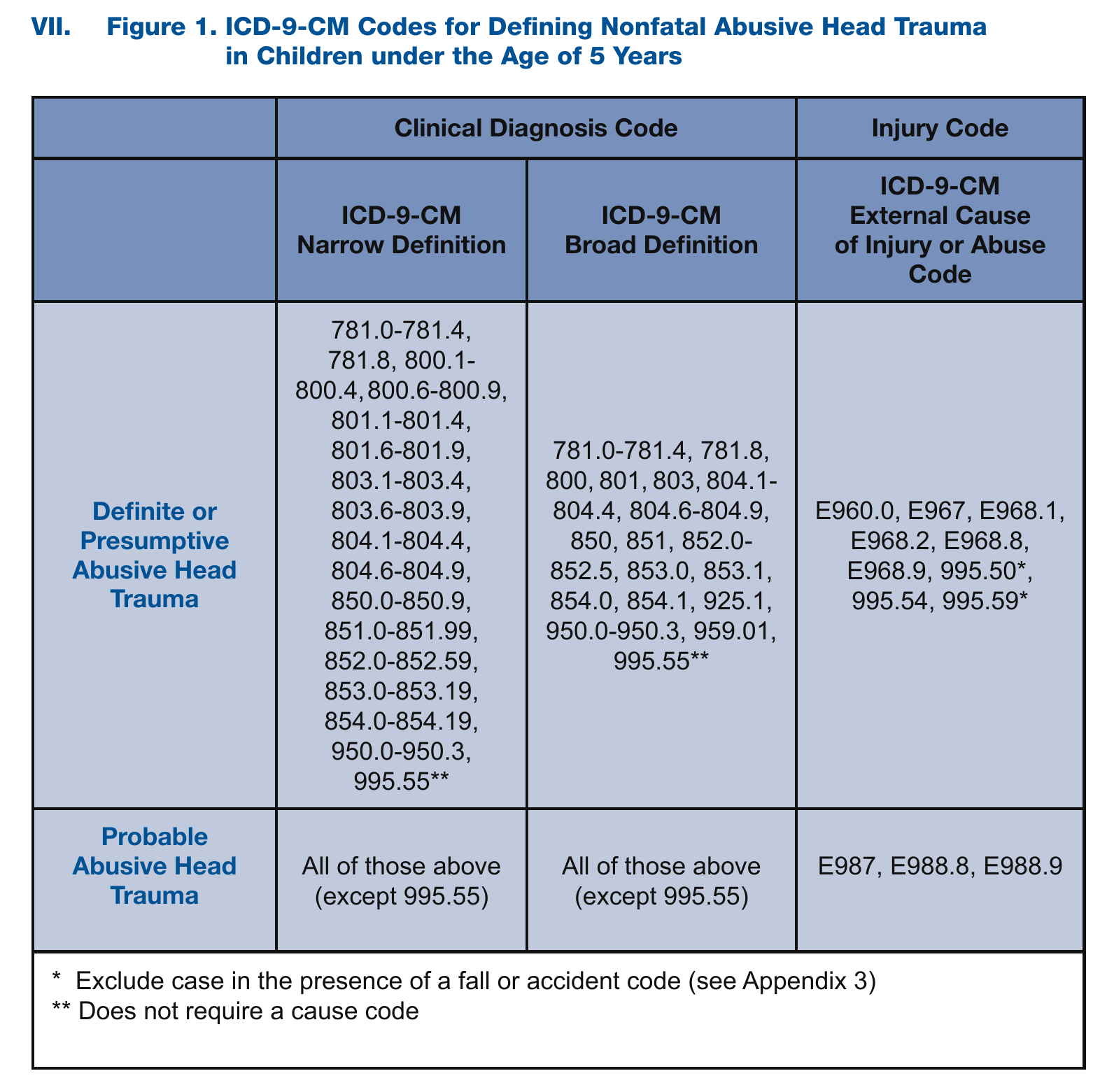} 
   \caption{Table from \cite{parks_2012_1} showing the diagnosis and injury codes for diagnosing AHT in nonfatal cases, allowing for a definite/presumptive diagnosis and a probable diagnosis.}
   \label{fig:parksfig1}
\end{figure}

\begin{figure}[!ht]
   \centering
   \includegraphics[width=.8\textwidth]{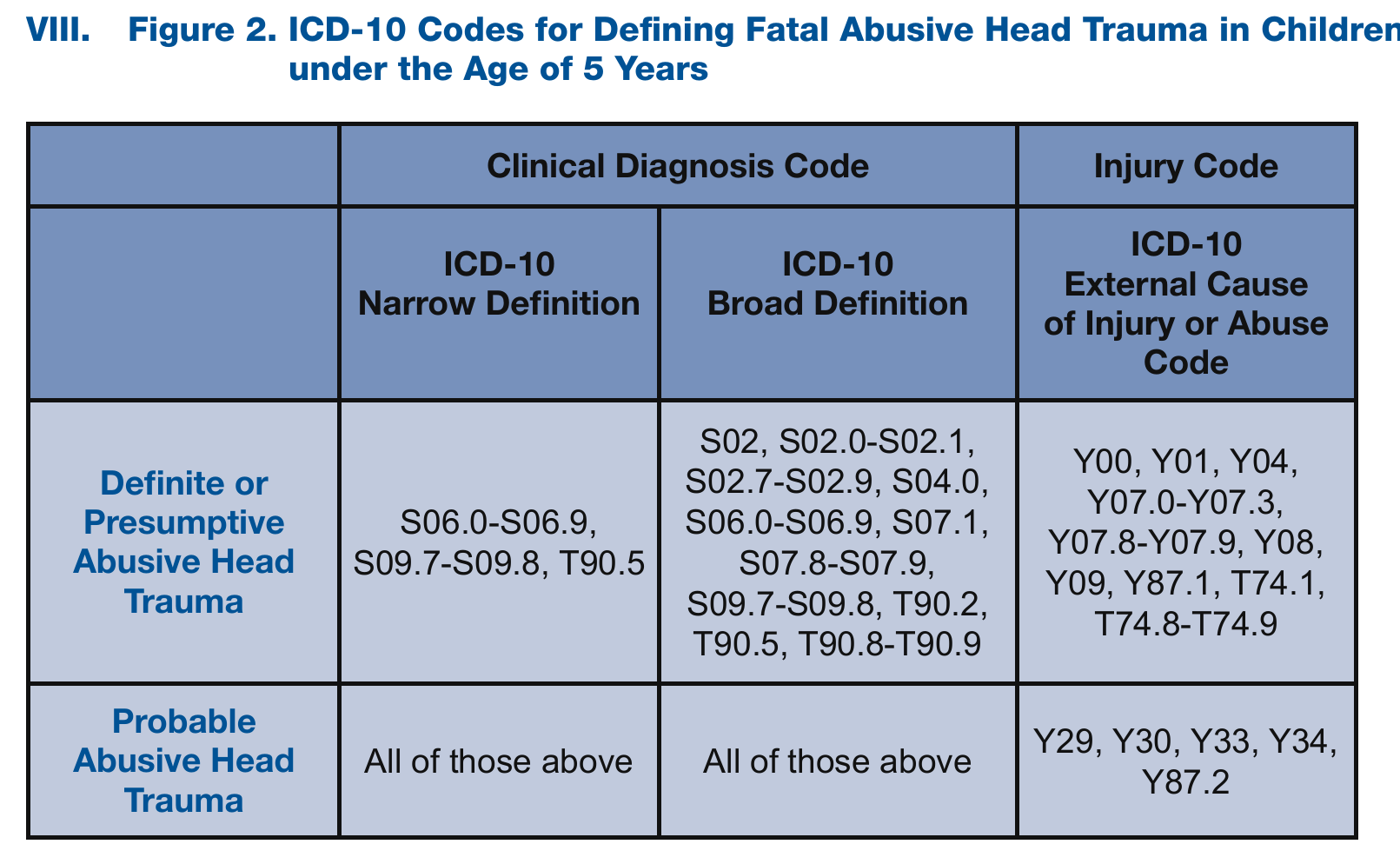} 
   \caption{Table from \cite{parks_2012_1} showing the diagnosis and injury codes for diagnosing AHT in fatal cases, allowing for a definite/presumptive diagnosis and a probable diagnosis.}
   \label{fig:parksfig2}
\end{figure}

The CDC is aware of the heterogeneity of SBS/AHT diagnoses. It has attempted to standardize diagnostic criteria by using multiple medical (ICD) codes and categorizing cases as definite, presumptive, or probable AHT. In March 2008, the CDC organized a workshop of experts \citep{cdcworkshoponsbs}, including pediatricians, child maltreatment experts, AHT experts, coding experts and experienced state health department personnel, to develop code-based case definitions for both non-fatal AHT, applicable primarily to hospital discharge data (see Figure \ref{fig:parksfig1}), and fatal AHT, applicable primarily to death certificate data (see Figure \ref{fig:parksfig2}). The panel reached a consensus on case definitions, allowing for the certainty of a diagnosis. 

However, the fact that there are 23 pages of ICD codes that are included in the SBS/AHT diagnosis, begs the question of whether the CDC's recommendation is a clearly defined standard for diagnosing SBS/AHT. While this type of approach is usually helpful to standardize diagnostic criteria for a condition, there could be problems with using it for SBS/AHT \citep{chp16epidemiology}. First, there is no explanation in the CDC report about whether the workshop organizers invited physicians who have different opinions about the way SBS/AHT has been diagnosed. The selection of physicians could change the diagnostic criteria dramatically. Second, the diagnostic criteria selected by the CDC does not include any information about the context that the physician used to make the diagnosis. Did the physician use information about the defendant, from the police, or a nurse's opinion, to determine the child was abused? This needs to be noted because it might be a piece of task-irrelevant contextual information \citep{cuellar2022probabilistic}. Finally, the circularity already explained could be a problem: If physicians who have diagnosed SBS/AHT, after observing certain signs, declare that children with those signs should be diagnosed with SBS/AHT, this is simply a circular way to say that they have been correct all along -- i.e., there is no external validation showing that they are correct.

Furthermore, the American Academy of Pediatrics recommended in 2012 that pediatricians use the term ``Abusive Head Trauma'' rather than ``Shaken Baby Syndrome'' to encompass various injury mechanisms \citep{findley2011shaken}. Nonetheless, both terms continue to be used interchangeably in medical and legal settings, adding to the confusion and inconsistency in diagnosing AHT. 

Some physicians believe that the CDC definition needs revision following new evidence, such as the report from \cite{swedishreport}, also known as the Swedish report, which challenges the shaking hypothesis. The Swedish report reviewed the scientific literature on SBS and concluded that the evidence supporting the diagnosis was weaker than previously thought. The report pointed to studies that showed how symptoms associated with SBS/AHT could result from conditions like neurological conditions or genetic diseases. The report's findings have fueled a broader reevaluation of the diagnostic criteria used by pediatricians and forensic pathologists, leading some to question whether SBS/AHT has been overdiagnosed in recent years.

\subsection{Medical and legal decisions are combined}

The reliance on medical diagnoses in SBS/AHT cases raises significant ethical and legal concerns. One of the most contentious issues is the blurring of the lines between medical diagnosis and legal judgment. When a physician testifies that a child has suffered from ``abusive head trauma,'' they are not merely describing a medical condition---they are making an implicit claim about the intent of the caregiver. This can lead to wrongful convictions if the diagnosis is based on flawed or incomplete evidence. Furthermore, a critical issue is that the subjective nature of SBS/AHT diagnoses can lead to a high degree of variability in how cases are handled legally. Inconsistent diagnoses can result in unequal treatment under the law, where the outcome of a case might depend heavily on which physician or medical team performs the diagnosis. This variability undermines the fairness of legal proceedings and can lead to significant injustices.

SBS/AHT is unique in that other diagnoses, such as sexual abuse, also intersect with legal decisions, but SBS/AHT cases involve very young children who cannot testify, making the physician's opinion pivotal. This intuition-based approach taken by the physicians, while informed by clinical experience, lacks the scientific testing for reliable diagnoses. 

Furthermore, the physician does not get feedback about whether his or her previous opinions were correct. Thus, the reliability of their diagnoses cannot be established by experience-based learning, since they do not learn whether they were correct and so experience does not offer sufficient guarantees of reliability. That's why, for example, other medical opinions are admissible, despite empirically based error rates, because those opinions arise in a context where experience actually does provide some assurances of reliability--where, for example, the physician can learn from experience because the physician gets feedback (in the form primarily of patient response to diagnosis and treatment). Some \citep{findley2011shaken} have argued that SBS/AHT is different in that regard, because there is no feedback and hence no opportunity to learn from experience.

Once the medical decision enters the legal realm, it must follow the legal admissibility standards for scientific statements. For instance, in a state with the Daubert standard \citep{daubert}, the trial court must consider the following factors to determine whether the expert's \textit{methodology} is valid: 1) Whether the technique or theory in question can be, and has been tested; 2) Whether it has been subjected to publication and peer review; 3) Its known or potential error rate; 4) The existence and maintenance of standards controlling its operation; and 5) Whether it has attracted widespread acceptance within a relevant scientific community. In this case, the methodology is whatever methods the physicians (or medical examiner/coroner) uses to determine whether a child has SBS/AHT. The Frye standard focuses only on point 5.

Regarding the five Daubert requirements, 1) The diagnosis of SBS/AHT has not been tested. Whether it \textit{can} is an unanswered question. To test it, physicians could participate in a black box study in which they are given case studies and they go through the process of making a diagnosis. Then their results are compared to the researcher's known ground truth. The problem with this is that it would be difficult to generate cases, since we do not know with certainty what happens when a child is shaken, and it would be difficult to recreate all the contextual information that a physician uses to make their diagnosis. Nevertheless, a black box study could be achieved with certain simplifications of reality, and then the remaining question would be whether the test is realistic enough to be relevant to real casework. 2) The results of this test would yield publications in peer-reviewed journals and 3) error rates. 4) As aforementioned, the way physicians make their diagnoses is heterogeneous. 5) Whether SBS/AHT has widespread acceptance in the relevant scientific community is unclear but doubtful given that there is a heated controversy, as described in the introduction. Nevertheless, the National Center on Shaken Baby Syndrome claims that there is a consensus about the diagnosis according to a number of medical organizations. The statement is published in Pediatric Radiology \citep{choudhary2018consensus}. \cite{narang2016acceptance} even describes that SBS/AHT are ``still generally accepted by physicians who frequently encounter suspected child abuse cases, and are considered likely sources of subdural hematoma, severe retinal hemorrhages, and coma or death in young children.'' The dada analyzed in this article, however, shows that several professions, like pathologists, are not unanimous about their views on SBS/AHT. 

The Federal Rules of Evidence 702 (Fed. R. Evid. 702) states that ``A witness who is qualified as an expert by knowledge, skill, experience, training, or education may testify in the form of an opinion or otherwise if the proponent demonstrates to the court that it is more likely than not that: (a) the expert's scientific, technical, or other specialized knowledge will help the trier of fact to understand the evidence or to determine a fact in issue; (b) the testimony is based on sufficient facts or data; (c) the testimony is the product of reliable principles and methods; and (d) the expert's opinion reflects a reliable application of the principles and methods to the facts of the case.

Regarding the four FRE 702 requirements, points a) and b) are for the trier of fact to determine. Point c) requires that the principles and methods, i.e., making a SBS/AHT diagnosis, have been shown to be reliable. In other words, that they have been tested empirically to have low error rates. Point d) requires that the physician actually applied the principles properly in a specific case, and this depends on the case. 

The PCAST 2016 report, inspired by the FRE 702, describes the concepts of ``foundational validity'' and ``validity as applied''. SBS/AHT does not have foundational validity because it has not been shown to be accurate, repeatable, and reproducible with empirical and well-designed studies. For it to be valid as applied, PCAST recommends that an uncertainty measure be given along with the scientific determination. For instance, in firearms, PCAST states that the expert should report the ``overall false-positive rate and sensitivity for the method established in the studies of foundational validity.'' And the expert should not make claims or implications that go beyond the empirical evidence and the applications of valid statistical principles to that evidence. If there are no studies demonstrating foundational validity, validity as applied cannot exist either.

Some might argue that the intuition-based approach of the physicians can be recognized as admissible under the rules of evidence, if it is considered ``clinical expertise'' based on experience, not on epidemiological studies. What is the scope of clinical expertise, and how it can bypass the requirements of scientific foundational validity, is left as an open question worth studying further. Others might disagree with these responses to the legal requirements for the admissibility of scientific statements, for instance the \cite{prosecutorguidelines}.

In Michigan, a court judgment \citep{michigancase} emphasized the impropriety of physicians diagnosing injuries as ``abusive head trauma'' or child abuse, as these terms imply intent or moral culpability, which are legal determinations. The ruling highlighted the need for clearer boundaries between medical opinion and legal conclusions, urging that medical experts should provide objective evidence without making inferences about criminal intent.

Remarkably, the author of the first article on SBS/AHT, Norman Guthkelch \citep{guthkelch1971} regretted having written that article in 1971 because it combined medical and legal determinations. He recommended that the SBS/AHT diagnosis be renamed to exclude any reference to the cause of the condition, as ``Infant retino-dural hemorrhage with minimal external injury.'' \citep{guthkelch2011problems}
Separating the medical diagnosis from the legal determination of abuse clarifies the fact that the physician's task is to make a medical determination based on physical and observable clinical signs and symptoms, and the physician should be considered epistemologically unable to assess whether the defendant abused a child. The trier of fact's task is to make a legal determination about whether there was child abuse, as he or she has the epistemic ability \citep{cheng2022consensus}, i.e., ubiquitous expertise, to determine whether the child was abused. This separation would indeed clarify the validity of the medical diagnosis, and it is necessary because the legal implications of a SBS/AHT diagnosis are profound. In criminal cases, the diagnosis can lead to severe penalties, including long-term imprisonment or even the death penalty. The stakes are high, and the potential for miscarriages of justice is significant.

\subsection{Fatal cases have additional complications}

Fatal cases have a additional complications, since  the medical examiner or coroner (i.e., the pathology expert) must determine the cause of death and manner of death of the child. This requires an additional dataset. It also suffers from the same problem as mentioned in the previous section, in which the physician acts as the trier of fact when he or she determines that there was abuse. The task of the pathologist is to determine the manner of death, and thus the contextual information is required for his or her task. The pathologist thus has the same task as the trier of fact. This is problematic, and it is an issue that is currently being discussed at the National Academies of Science by a study called ``Advancing the Field of Forensic Pathology: Lesson Learned from Death in Custody Investigations'' \footnote{\url{https://www.nationalacademies.org/our-work/advancing-the-field-of-forensic-pathology-lesson-learned-from-death-in-custody-investigations} Accessed 11-19-2024}. In addition, as mentioned in Section \ref{sec:contextualbias}, there is evidence that pathology experts suffer from contextual bias, which further leads to errors in the determination of abuse.

\section{Data about SBS/AHT}

\subsection{Available data does not contain the information necessary to understand current SBS/AHT diagnoses}

Numerous databases offer insights into injury-related healthcare data, but estimating the prevalence of Shaken Baby Syndrome (SBS) and Abusive Head Trauma (AHT) remains challenging due to fragmented data and inconsistent diagnostic criteria. Despite these obstacles, resources like the Kids' Inpatient Database (KID) \citep{kids} and Nationwide Inpatient Sample (NIS) \citep{nis} provide regional and national estimates using ICD coding systems to track diagnoses and external injury causes. However, variability in coding practices limits their reliability.

Emergency department data and inpatient discharge datasets, such as those from HCUP, complement each other by capturing less severe and more serious injuries, respectively. Regional resources like California's EPICenter \citep{epicenter} and national systems like the National Vital Statistics System (NVSS) \citep{nvss} focus on localized and fatal cases, though variability in classification persists. Privately compiled datasets also provide unique insights but are prone to selection bias.

According to CDC data \citep{parks_2012_1, parks_2012_2}, annual non-fatal AHT hospitalizations are estimated at 10,555, with significant lifelong consequences for survivors. Incidence rates of SBS/AHT in children under one year range from 33 to 38 per 100,000. Prevention programs like the Period of PURPLE Crying aim to reduce these numbers, though trends show mixed outcomes. Legal data is similarly fragmented, complicating efforts to track prosecutions and convictions. For example, in the UK, confidentiality in Family Court proceedings obscures judicial outcomes, further limiting transparency in SBS/AHT cases.

Regarding legal data, assessing the number of SBS/AHT cases that lead to prosecution and conviction is also problematic due to the fragmented nature of legal records. Records are maintained separately across jurisdictions, often in inconsistent formats, complicating efforts to obtain a comprehensive view. Efforts to unify these records, such as the Criminal Justice Administrative Records System (CJARS) \citep{cjars}, have made limited progress. This fragmentation hinders the ability to determine the true scope of legal outcomes associated with SBS/AHT cases, including the number of prosecutions, convictions, plea deals, and sentences.

This issue extends beyond the United States. In the UK, the majority of SBS hearings are held in Family courts, which are confidential, making case numbers, outcomes, and evidence used largely unknown. Researches note that less than 5\% of decisions are public, resulting in a lack of scrutiny and transparency in these cases \citep{chp21ptolemy}. The secretive nature of these hearings further obscures the understanding of SBS/AHT prevalence and judicial outcomes.

Efforts to standardize data collection and unify criteria across datasets are critical for improving prevalence estimates, understanding trends, and addressing diagnostic and legal challenges.

\subsection{How data could help us understand what is happening with SBS/AHT diagnoses}

The way that data is collected about SBS/AHT currently only includes medical information, and thus the crucial information used to make the diagnosis (e.g., family interviews, police interviews, multidisciplinary team's opinion, etc.) is not visible in the data. This means we cannot understand the diagnostic process by only using the available data. 

\begin{figure}[!ht]
    \centering
    \includegraphics[width=\linewidth]{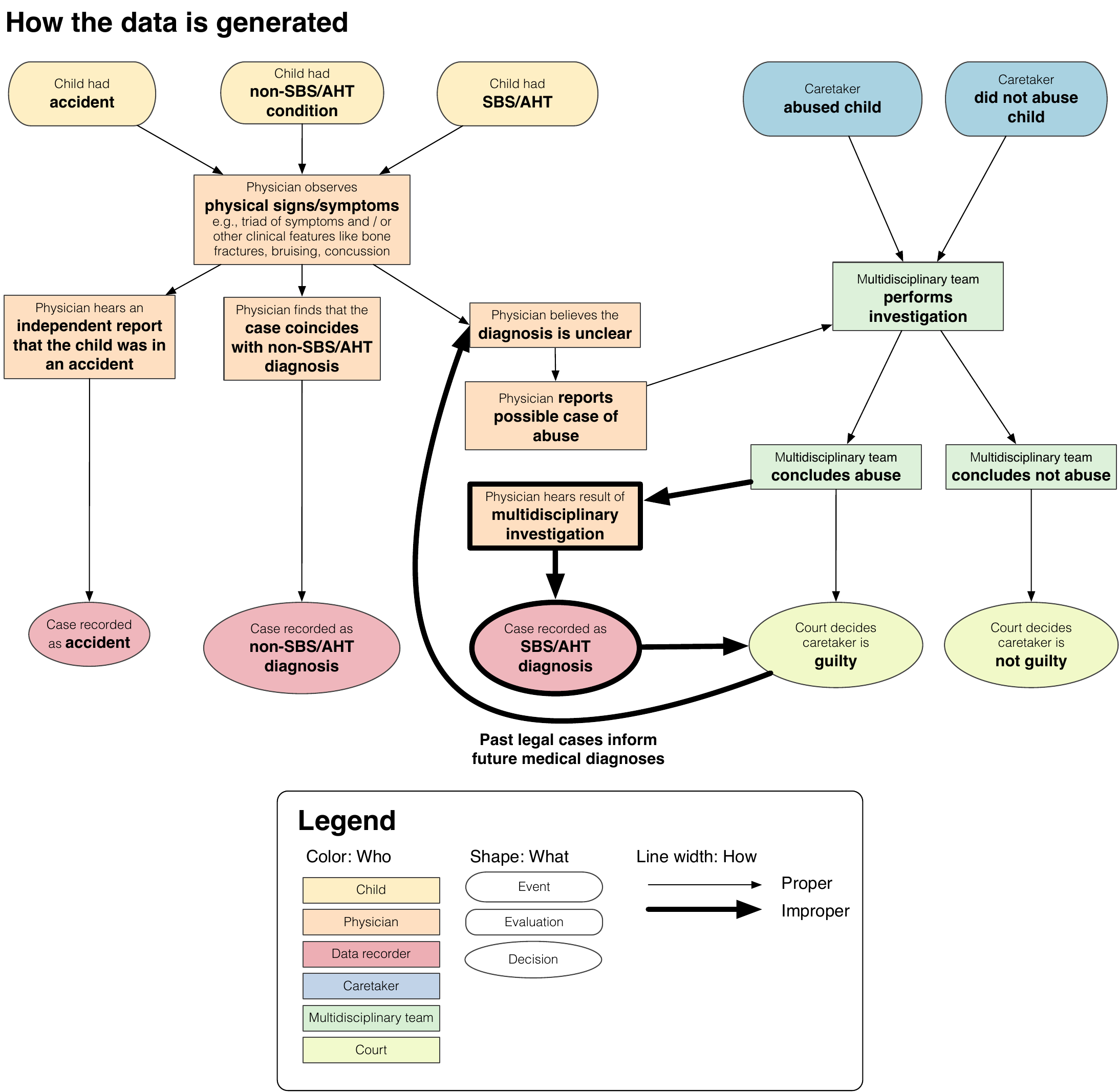}
    \caption{Graph depicting a hypothesized data generation process for current practice, simplified for clarity. The dark arrows and frames depict the improper reasoning currently used in diagnosing SBS/AHT. To see how the data \textit{should be} generated, see Figure \ref{fig:graph-shouldbe} in the Appendix, which excludes all the pieces with thick lines. (For more information about this Figure, see Footnote \ref{footnote:graph-is}.)}
    \label{fig:graph-is}
\end{figure}

This process is depicted in Figure \ref{fig:graph-is}.\footnote{Some comments about Figure \ref{fig:graph-is}: i) When multiple arrows arrive at a node, this indicates that all cases lead to some version of that node. And when multiple arrows leave from that node, this indicates that some of the cases result in one or another of the subsequent nodes. 
ii) Independently witnessed abuse cases were not included because there has been no case of such a case \citep{findley2011shaken}. iii) The thick arrow from the Guilty verdict to Diagnosis is unclear indicates a circular feedback loop from past legal cases informing future medical diagnoses. Even though the physician might think the diagnosis is unclear, past legal cases give more weight to the possibility that this case was caused by abuse. iv) This graph does not include fatal cases, for simplicity. Those cases should stay within the legal side, and not enter back into the medical side. But, this is complicated because pathology requires a determination of manner of death, which combines medical and legal decisions by definition. \label{footnote:graph-is}} This figure shows a hypothesized data generation process for cases involving children with brain conditions. The aim is for the physician to classify each case as abuse, accident, or another medical condition, and for the courts to determine whether the caretaker is guilty of child abuse. However, many clinical features--such as retinal hemorrhage, cerebral edema, and subdural hemorrhage--are common across all three scenarios, making the classification challenging. To identify accidents, witness information (e.g., evidence that the child was in a car crash) is typically relied upon. For medical conditions other than SBS/AHT, differential diagnoses are used by matching the observed symptoms to those of known conditions. For the rest of the cases, which are unknown, contextual information from police investigations, court proceedings, and child protection teams often informs the decision. 

The challenge with diagnosing abuse lies in a problematic reasoning process. Physicians might hear the results of a multidisciplinary team's legal investigation, which is a subjective process that might have errors, and this might inform their medical opinion. This in turn leads to a legal determination of abuse. The problem here is the mixing of legal and medical decisions, which is especially problematic because there could be errors in the subjective determinations of the multidisciplinary team.

Then, in future cases, physicians may diagnose abuse based on the same symptoms and contextual factors they have historically associated with it. This creates a feedback loop: their conclusions are repeatedly reinforced without external validation, meaning they may never know whether these abuse diagnoses were accurate in the first place. 

The data should be generated by removing the processes with thick black arrows, as shown in Figure \ref{fig:graph-shouldbe} in the Appendix. In the setting showing how the data should be generated, the physician's last step should be to determine that the diagnosis is the child's physical signs/symptoms with unknown cause, and a suspicion of abuse. The suspicion of abuse could come from speaking with the family, who presumably have brought the child to the hospital. But any conclusion that includes the context should be left at most at a suspicion of abuse. Then, the case should go to the multidisciplinary team, which investigates the caretaker. The multidisciplinary team should not communicate back with the physician after the investigation, such as to not influence the physician's medical opinion. Any communication from the legal team back to the physician has the potential to produce contextual bias in future cases.

\newpage

\begin{figure}[!ht]
    \centering
    \includegraphics[width=\linewidth]{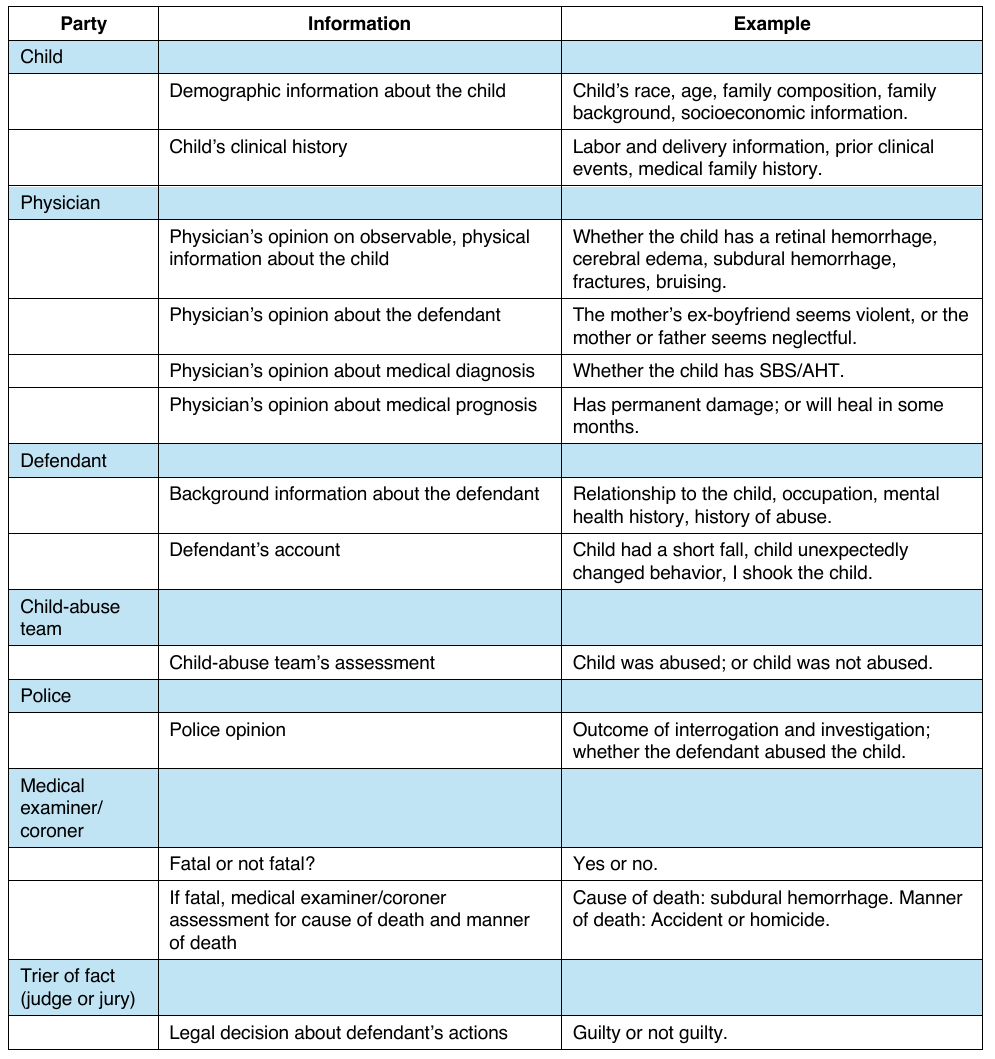}
    \caption{Example of what high-quality data to study SBS/AHT might look like. This includes information from the different actors' opinions (physician, defendant, child abuse team, the court, etc.) about whether a child was abused. Having all this information would give clarity to what happened medically and legally, and the best way to treat the child and the defendant.}
    \label{fig:idealdatatable}
\end{figure}

It can be useful exercise to imagine what a better dataset would look like for studying a condition, even if it is difficult to acquire. Figure \ref{fig:idealdatatable} shows a table with a list of pieces of information that could be useful to determine what happened, or at least how a child came to be considered abuse -- whether it was because of the child's demographic information, the defendant's background information, the police's opinion, the judge's opinion, etc.

This table is an attempt to clarify what information is objectively observed (e.g., retinal hemorrhage), what is subjectively decided (e.g., the nurse believes the mother is neglectful), and what is motivated by a legal decision (e.g., the child might have been abused by the defendant, and if this is the case, it's safer to convict the defendant so they do not continue abusing children). 

If we had this data, we could find trends about how it is decided that children are abused in different locations nationally and globally. Is it a decision that is mostly made by physicians' gut feelings? Or is it an attempt by the legal system to prevent more harm? Answering these questions could shed light on how best to treat the live children with this condition, and how best to convict individuals guilty of child abuse.


\section{Example of a data analysis with medical and legal data: The causal effects of the PURPLE program}\label{sec:exampleanalysis}

We seek to show an example of how an analysis might look with richer data that includes information about both the defendant and the child. Our goal for doing this data analysis is simply to show an exercise in analyzing legal and medical data about SBS/AHT cases jointly. We do not claim that these results are reliable.

\subsection{The data: A privately collected convenience sample, with medical and legal information for each case}

We use a private dataset compiled by Susan Weston\footnote{Contact information: susan@susancanthony.com}. Weston began gathering cases in 2000 after her daughter-in-law faced accusations with no supporting history of abuse. By 2006, she subscribed to Google alerts for terms like ``Shaken Baby Syndrome'' and later adapted to include ``Abusive Head Trauma'' (AHT) as terminology evolved. In 2009, she organized her findings--444 cases at that time--into FileMaker, a database that she continued updating as she accessed more sources, including WestLaw and her library's newsbank. By February 2023, Weston's dataset had expanded to 5,928 cases spanning from 1956 to 2021. Finally, the race data was limited, so we had an undergraduate student at the University of Pennsylvania look for information about the cases on the internet and record race for the infant and defendant if it was available explicitly or as an image. Note that this race assessment is unverified by a second source.

The present dataset has an advantage over other datasets: It addresses the medico-legal paradigm of SBS/AHT by merging health data with legal outcomes, allowing legal data to challenge the certainty and confidence in diagnosis. But, it has more important disadvantages: it is not a probability sample or a census. Instead it is a convenience sample collected by an individual based on media and word-of-mouth. Given that this dataset is composed of a subset of publicized cases, there is selection bias in that typically only serious cases (typically those leading to death) are covered by the media. Furthermore, similar to the limitations of other studies, we cannot be sure whether the publicized cases are under- or over- reported. Thus, we cannot trust any inferences performed with the data. Nevertheless, it is an interesting exercise to study what the publicly available cases of SBS/AHT are like.

\begin{figure}[!ht]
   \centering
   \includegraphics[width=.9\textwidth]{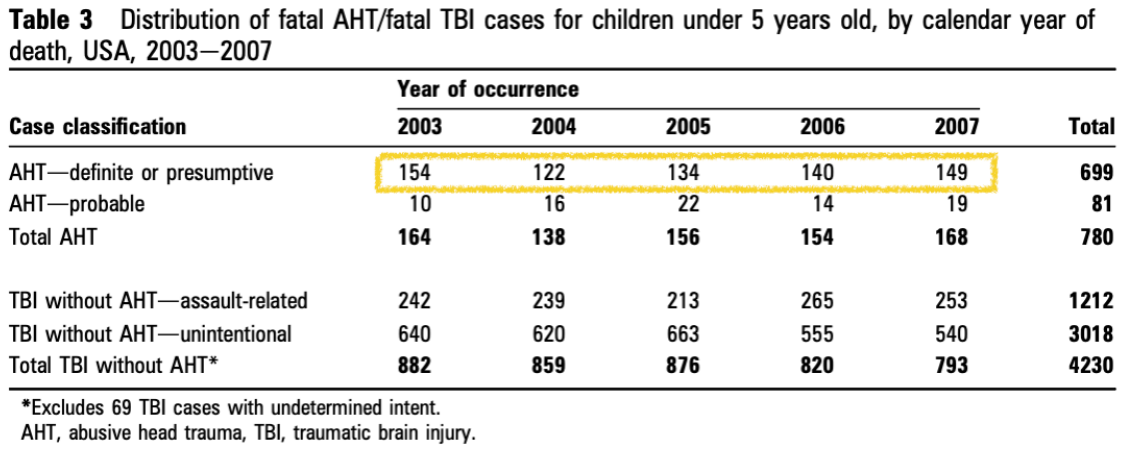} 
   \caption{Table from \cite{parks_2012_1} showing the number of diagnoses.}
   \label{fig:timedata1}
\end{figure}

\begin{figure}[!ht]
   \centering
   \includegraphics[width=.4\textwidth]{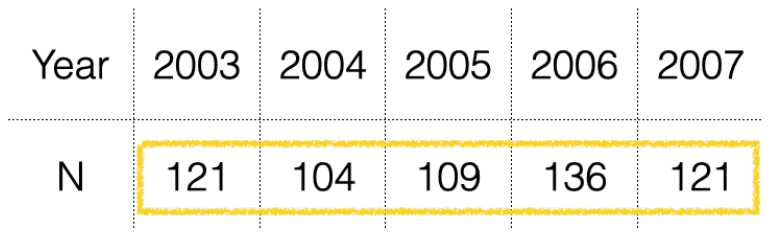} 
   \caption{Table from private dataset, showing the number of diagnoses.}
   \label{fig:timedata2}
\end{figure}

How does this private dataset compare to publicly available data? Figures \ref{fig:timedata1} and \ref{fig:timedata2} show the counts for fatal cases of AHT, comparing the data from \cite{parks_2012_1} and the private dataset. The fatal cases are likely to have less underreporting, as in the data from the Uniform Crime Reports \citep{blackman1984flaws}. This heuristic comparison shows that the numbers of fatal cases are similar across the two databases, and thus perhaps the underreporting in the private dataset is not so high -- or at least it is not much higher than that in the CDC data.

\subsection{Descriptive statistics}

\begin{figure}[!ht]
    \centering
    \includegraphics[width=.8\linewidth]{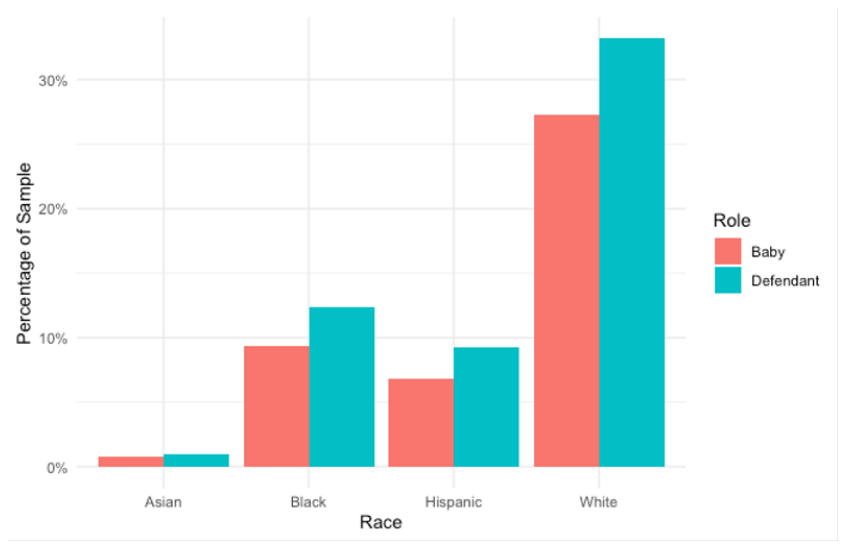}
    \caption{Racial Distributions of infants and defendants in the private dataset.}
    \label{fig:race}
\end{figure}

Summary statistics are shown in Figure \ref{fig:race} and Table \ref{tab:summarystats}. There is more race data about the defendants than the infants, but the racial distribution is similar for both. In terms of gender, the infants are not too imbalanced, but the defendants are more likely to be male. Regarding location, Figure \ref{fig:map} shows the distribution of cases in the dataset across the United States. Cases were heavily concentrated in Florida (382 cases; 6.5\%), California (332 cases; 5.6\%), Pennsylvania (329 cases; 5.6\%), Texas (324 cases; 5.5\%), and Ohio (310 cases; 5.3\%). 

We study two measures of interest: the total number of cases within each state in each year, and trial outcome. The trial outcome variable indicates the resulting decision from the initial trial. Of the 5,928 cases in the dataset, a trial outcome is available for 5,026 cases. Of the 20 potential variable outcomes, two are of particular importance: ``plea bargain'' (2,240, or 44.6\% of available cases) and ``convicted'' (1,205, or 24.0\% of available cases).

\begin{table}[!ht]
    \centering
    \caption{Summary statistics of cases by gender and race}
        \label{tab:summarystats}
    \begin{tabular}{lcccc}
        \toprule
        & \multicolumn{2}{c}{\textbf{Child}} & \multicolumn{2}{c}{\textbf{Defendant}} \\
        \cmidrule(lr){2-3} \cmidrule(lr){4-5}
        & \textbf{Count} & \textbf{Percentage} & \textbf{Count} & \textbf{Percentage} \\
        \midrule
        \textbf{Gender} \\
        \hspace{5mm} Female & 2359 & 40\% & 1477 & 25\% \\
        \hspace{5mm} Male & 3277 & 55\% & 4332 & 73\% \\
        \hspace{5mm} Unknown & 292 & 5\% & 119 & 2\% \\
        \midrule
        \textbf{Race} \\
        \hspace{5mm} White & 1954 & 33\% & 2381 & 40\% \\
        \hspace{5mm} Black & 669 & 11\% & 890 & 15\% \\
        \hspace{5mm} Hispanic & 491 & 8\% & 666 & 11\% \\
        \hspace{5mm} Asian & 58 & 1\% & 67 & 1\% \\
        \hspace{5mm} Unknown & 2756 & 46\% & 1924 & 32\% \\
        \midrule
        \textbf{Total cases} & 5928 & 100\% & 5928 & 100\% \\
        \bottomrule
    \end{tabular}
\end{table}

\subsection{The treatment: The Period of PURPLE Crying program}

This example investigates racial disparities in AHT diagnoses and criminal trials as well as to evaluate the Period of PURPLE Crying program within the United States. A private dataset that combines medical and legal information for 5,928 cases is examined. Further, year implementation data from the National Center on Shaken Baby Syndrome is utilized for a difference-in-differences analysis to evaluate the effectiveness of the PURPLE program. We find that the PURPLE program had no significant effect on the number of AHT cases nor on convictions. Analysis shows that the program increased plea bargains, although this was isolated 9 years post-implementation. An objective definition and criteria for diagnosing AHT is necessary. Until then, the PURPLE program should expand to other states given that it educates parents on infant crying patterns. 

The Period of PURPLE Crying program aims to teach parents about healthy crying characteristics in infants, including the age of peak crying, its unexpectedness, an infant's resistance to soothing efforts, a pain-like face, long-lasting duration, and the likelihood of increased crying in late afternoons and evenings \cite{ncsbs}. The program includes a health professional-led education session and a combination of educational brochures and DVDs. Figure \ref{fig:map} below depicts the current status of PURPLE programs across each state. 

\begin{figure}[!ht]
   \centering
   \includegraphics[width=\textwidth]{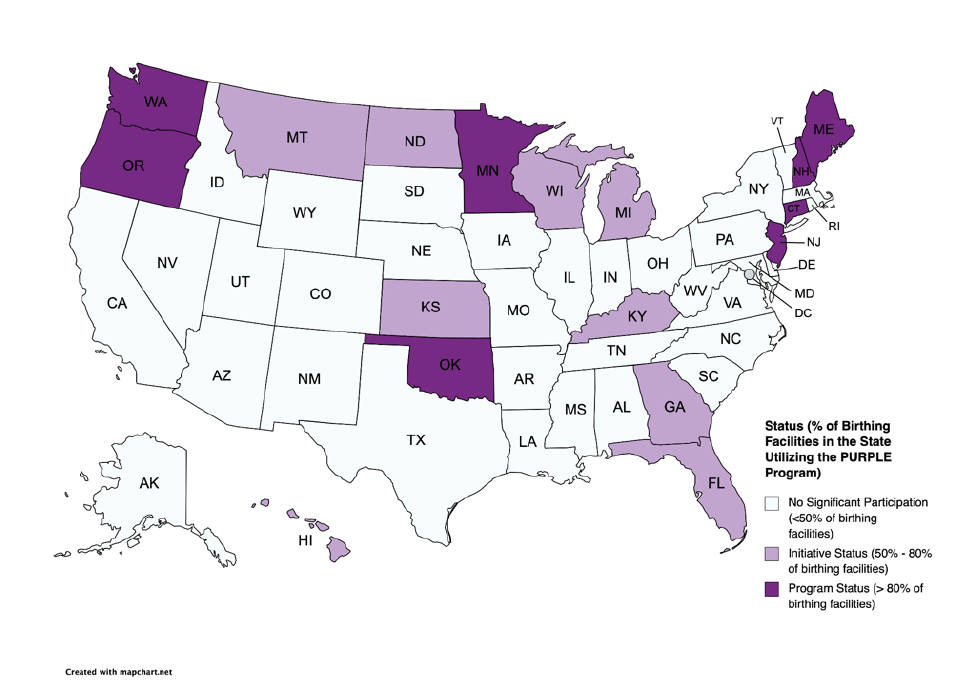} 
   \caption{Status of the Period of PURPLE Crying Program across the United States. This map was created using information received from the campaign manager at the National Center on Shaken Baby Syndrome in February 2023. The ``initiative status'' is given to states for which 50\% to 80\% of their birthing facilities have implemented the program. Once 80\% or more birthing facilities across the entire state implement the program, the state is given ``program'' status. The majority of states did not have enough participation across their birthing facilities to be considered initiatives or programs.}
   \label{fig:map}
\end{figure}

Overall, similar programs seem to show promise, despite the PURPLE program itself not having had a significant impact on AHT/SBS incidence. Regardless of the results being inconclusive, research suggests that a \$5 investment per child for the program can result in nearly \$300 of avoided cost by society and the healthcare system collectively \citep{beaulieu2019lifetime}. Thus, given the debate over the accuracy of AHT/SBS diagnoses, more research needs to be carried out before ruling out the PURPLE program as an effective intervention all together.

\subsection{Method: Difference-in-differences with staggered adoption}

In the standard Difference-in-Differences (DiD) setup, there are two time periods and two groups: during the first period, no units receive treatment, while in the second period, some units are treated (the treated group) and others are not (the comparison group). Under the assumption of parallel trends--meaning that, without treatment, the average outcomes for both groups would have evolved similarly over time--the average treatment effect for the treated (ATT) can be calculated by comparing the change in outcomes for the treated group to the change in outcomes for the comparison group.

Since the period PURPLE program was implemented in different states at different times, the setup here is more complex than the standard DiD approach. \cite{callaway2021difference} introduces a unified framework for analyzing such complex setups, focusing on identifying, aggregating, and estimating causal parameters while accommodating treatment effect heterogeneity and dynamic effects. The approach emphasizes the group-time average treatment effect -- a flexible and interpretable parameter defined by when units are first treated -- offering advantages over standard two-way fixed effects regressions by improving transparency, objectivity, and flexibility in causal analysis. 

Our analysis involves multiple treatment periods, with staggered adoption of the PURPLE program across treatment states. Further, instead of a singular treatment group, there are multiple groups, with each state being grouped by their year of implementation. Thus, the \cite{callaway2021difference} method, with its accompanying package in the R software (\texttt{did}), is a good fit for this dataset. The package allows for DiD analysis in situations where treatment groups experienced the intervention at different times. DiD designs in which treatment groups receive their treatment at different times are called event studies. For the event study plots used in this analysis, data was aggregated such that each treatment state received the intervention at time 0, centering the remaining data based on years before or after treatment . Please refer to Table 2 for the list of treatment states and the year that they first joined the program. 

The difference-in-differences models used in this analysis are,
\begin{equation}
N_{it}= \alpha_i+ \lambda_t+\Sigma_{\tau=-q}^{-1}\gamma_\tau D_{s\tau}+ \Sigma_{\tau=0}^m \delta_\tau D_{s\tau}  \epsilon_{ \sigma \tau},
\end{equation}
where $N$ refers to the number in question (SBS/AHT cases, convictions, and plea bargains, for each of three models, respectively). $i$ indexes the state, $t$ indexes the year, the treatment occurs in year zero, and we include $q$ leads and $m$ lags. In the study event plot, the intervals before the treatment are represented by $\gamma_\tau$ and the intervals after treatment are represented by $\delta_\tau$. There is possible heterogeneity in the treatments between states, and this might lead to bias in the results, a problem referred to as the ``Bacon decomposition'' \citep{nichols2019bacon}. 

In our results, we do not include the tables of estimated coefficients for these models purposefully, because we do not believe that these models are correct due to the data being incomplete. In other words, since the private dataset was collected by one individual by following the media, it is simply not reliable as a source of information about the effectiveness of a government program.

\subsection{Results}

\begin{figure}[!ht]
    \centering
    \includegraphics[width=.9\linewidth]{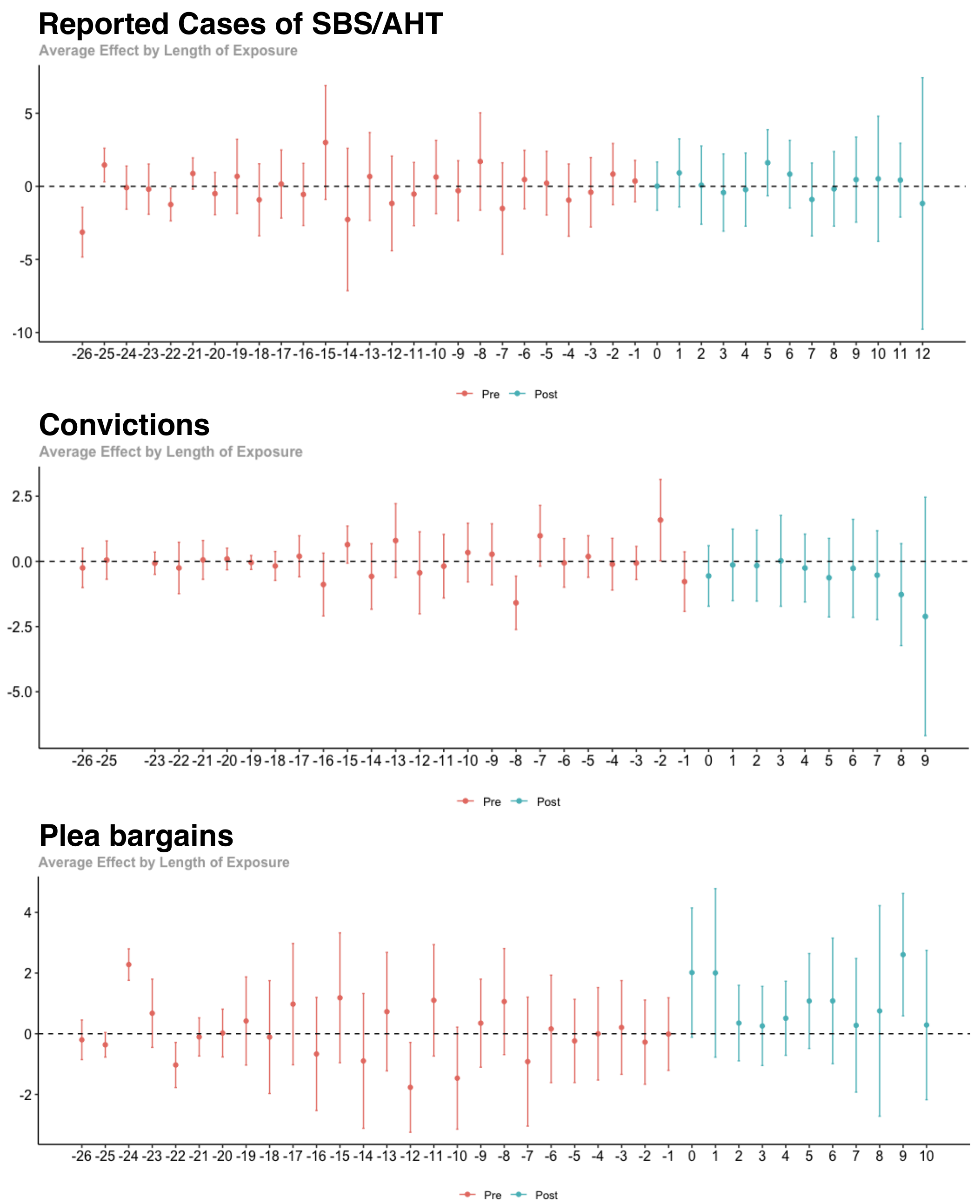}
    \caption{Event-study plots for the three models fit, from 1985 to 2021. The blue post-implementation period shows that there was no significant causal impact of the period PURPLE program on the three outcomes.}
    \label{fig:results}
\end{figure}

The event study plots in Figures \ref{fig:results} reveal no statistically significant effect of the Period of PURPLE Crying program on the number of SBS/AHT cases, convictions, or plea bargains. Before implementation, trends were largely parallel for both cases and convictions, as confidence intervals (CIs) mostly included zero. Post-implementation, CIs for SBS/AHT cases and convictions continued to cover zero, indicating no causal impact of the program. 

Nine years after implementation, plea bargains did increase, likely influenced by child abuse physicians testifying about the program's warnings, which may have pressured some defendants to accept plea deals. Limitations of our dataset underscore the need for broader data collection, including comparisons to electronic health records, to better assess the program's impact on SBS/AHT trends. Recall, however, that we do not endorse these inferential results as accurate because the data is a convenience sample.

\section{Conclusion}

This article advocates for a comprehensive effort by researchers to collect better data (as described in Figure \ref{fig:idealdatatable}) to understand how diagnoses are made. Better data could also help bring clarity to the foundational validity (accuracy, repeatability, reproducibility) of the SBS/AHT diagnosis, which might help practitioners report their conclusions with a measure of uncertainty, a practice that would help triers of fact (and the society at large) know how much trust to place in these decisions. Medically, understanding how accurate the diagnosis is can help improve treatment outcomes. Legally, this can help prevent wrongful convictions and miscarriages of justice. The role of physicians should be limited to physical observations, with reported measures of uncertainty, and legal determinations of abuse should be left to the trier of fact. Addressing these issues is crucial to safeguard the rights of accused individuals, to protect vulnerable children from abuse, and to provide children with appropriate treatment for their condition. 

An ideal source of data would be one that contains ground truth information about abuse. This includes cases of children who have the brain conditions currently attributed to SBS/AHT, some of whom were not abused and some of whom were abused. The key element is that the abuse needs to be independently verified, e.g., filmed or witnessed by a reliable individual or group. Independent verification cannot be replaced with confessions (there are unknown numbers of false confessions), legal determinations (there are unknown numbers of wrongful convictions), or the opinion of a multidisciplinary team (there are unknown numbers of incorrect determinations of abuse). Only with ground truth information about abuse can data be used to predict the cause of a child's brain condition.

Given that ground truth is not available, we cannot assess the accuracy of SBS/AHT diagnoses, and the data on SBS/AHT should not be used to design statistical or machine-learning models to predict abuse. Simply using statistical models does not eliminate the fact that the ground truth is missing. 

Since we cannot know how accurate the diagnosis is, this suggests that the SBS/AHT diagnosis should not be made at all by physicians. Instead, physicians should report an unknown cause for the brain condition and should report suspected child abuse to a legal party (e.g., the child abuse team or the police). Figure \ref{fig:graph-shouldbe} depicts how this decision should happen. 

However, data could be used to study how SBS/AHT diagnoses are made today. If the data included all the contextual information relevant to the case from the different parties involved, then statistical models or algorithms (like \cite{maguire2}'s) could be used to understand what types of information are being used to make a diagnosis and arrive at a legal determination. For example, if a physician used legal information to make a determination of abuse, which would be improper, then this could be revealed by the data. Models fit to this contextual data should be descriptive rather than prescriptive, since they could help researchers inside and outside the debate understand how decisions are being made, by whom, and at what point in the process. The example from Section \ref{sec:exampleanalysis} shows how more complete data could be used to understand more about medical diagnoses and legal decisions. Understanding how diagnoses of SBS/AHT are currently being made should be a priority in the research about SBS/AHT. 

The case of Robert Roberson, whose execution was stayed in October 2024 and is still unresolved at the time of writing, highlights the contentious debate surrounding the diagnosis of Shaken Baby Syndrome/Abusive Head Trauma. This ongoing controversy among researchers, physicians, pathologists, and legal professionals underscores the urgent need for objective, high-quality research on the topic. 

Some may argue that this article advocates for perfection or is overly idealistic. However, the recommendations offered here are far from unattainable. The solution lies in gathering better data and treating this diagnosis as any other scientific argument within the legal system. It should be the role of the trier of fact to determine whether a child with a brain condition was abused. Instead of including 23 pages of ICD codes as part of the SBS/AHT diagnosis \citep{parks2012pediatric}, physicians, medical examiners, coroners, and other relevant parties should focus on presenting evidence about observable signs, symptoms, as well as task-relevant contextual information, allowing the trier of fact to make an informed determination of guilt. Indeed, a discussion of what is task-relevant and task-irrelevant for a medical determination, perhaps organized by an entity such as the CDC, is needed.

Developing a better, evidence-based framework for diagnosing SBS/AHT could result in more accurate medical evaluations, fairer legal processes, and better outcomes for children and caregivers alike. Achieving this will require collaboration among medical professionals, legal experts, and researchers, which is a difficult but achievable goal that should be attempted in the pursuit of improving medical pediatric practices and the criminal justice system.

\bibliographystyle{abbrvnat}
\bibliography{sbsletterbib.bib}

\appendix

\section{Appendix}

\begin{figure}[!ht]
    \centering
    \includegraphics[width=.9\linewidth]{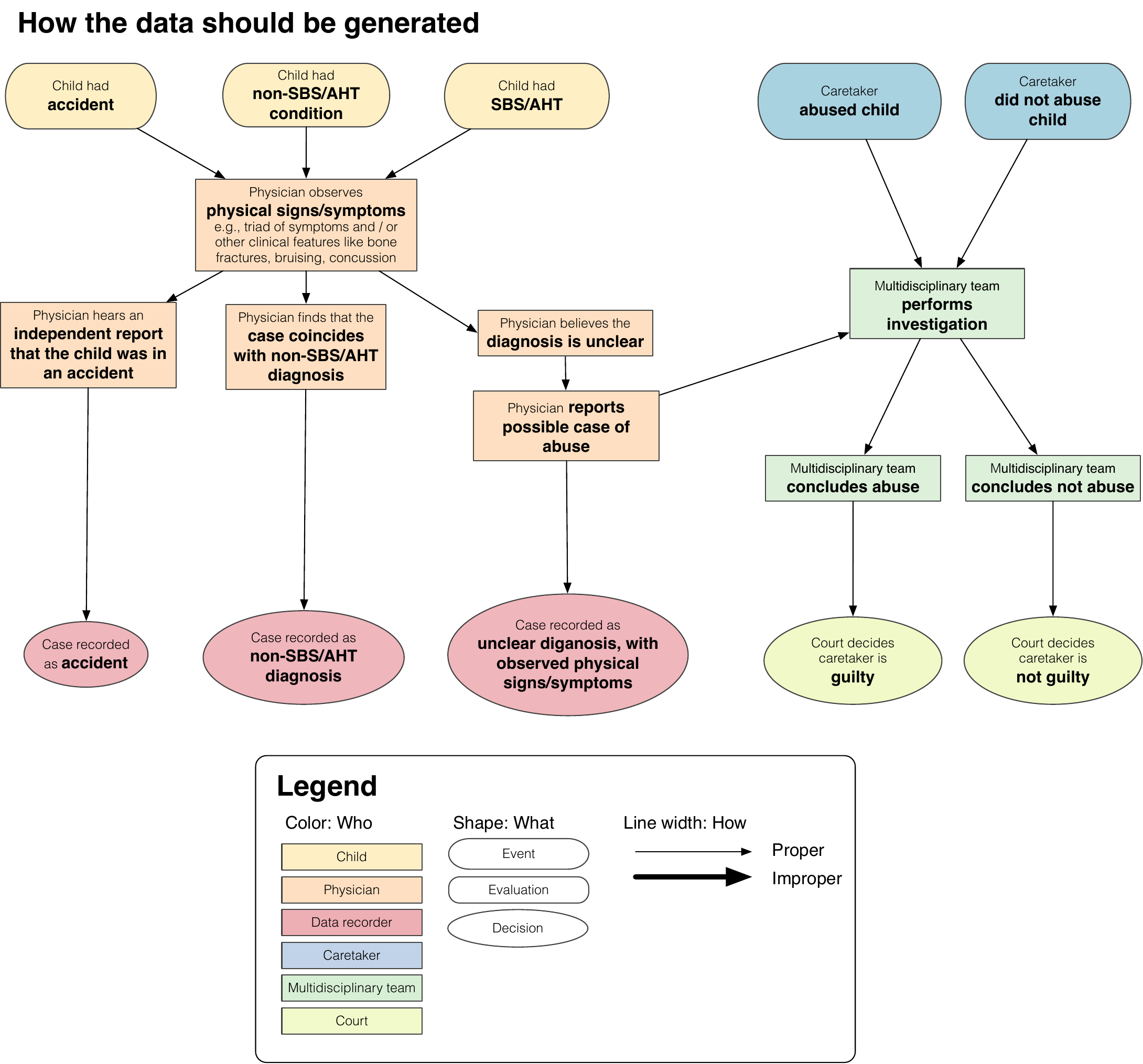}
    \caption{Graph depicting a hypothesized data generation process for how the process should be. This is the same as Figure \ref{fig:graph-is}, but with all the improper transfers of information removed.}
    \label{fig:graph-shouldbe}
\end{figure}

The data should be generated as in Figure \ref{fig:graph-shouldbe}.

\end{document}